\documentclass[a4paper,11pt]{article}
\pdfoutput=1
\usepackage[english]{babel}
\usepackage{jheppub}
\usepackage{amsmath,amssymb,graphicx,textcomp,enumerate,xspace}
\usepackage{soul} 

\usepackage[utf8]{inputenc}
\usepackage{fancyvrb} 
\VerbatimFootnotes 

\newcommand{\kt}{\ensuremath{k_{\rm\sss T}}\xspace}
\newcommand\sss{\mathchoice%
{\displaystyle}%
{\scriptstyle}%
{\scriptscriptstyle}%
{\scriptscriptstyle}%
}

\newcommand{\ttbar}{\ensuremath{t \bar t}\xspace}

\def\beq{\begin{equation}}
\def\beqn{\begin{eqnarray}}
\def\eeq{\end{equation}}
\def\eeqn{\end{eqnarray}}

\def\lq{\left[} 
\def\rq{\right]}

\def\({\left(} 
\def\){\right)}

\newcommand\as{\alpha_{\sss\rm S}}

\newcount\minutes 
\newcount\scratch 
\def\timestamp{%
\scratch=\time 
\divide\scratch by 60 
\edef\hours{\the\scratch} 
\multiply\scratch by 60 
\minutes=\time 
\advance\minutes by -\scratch 
---$\,$\hours:\null 
\ifnum\minutes< 10 0\fi 
\the\minutes}

\definecolor{mygray}{gray}{0.5}

\newcommand\pt{p_{\sss\rm T}}

\def\beeq{\begin{eqnarray}} 
\def\eeeq{\end{eqnarray}} 
\def\to{\rightarrow}

\newcommand\HCCF{\lq \HCCF H^{F} C_1 C_2 \rq_{c\bar{c};\,ab}}

\newcommand\plugin{MG5aMC-PWG}
\newcommand\pluginpy{MG5aMC\_PWG}

\newcommand\xDphijj{$\Delta\phi_{\sss j_1 j_2}$}
\newcommand\xDphiorjj{$\Delta\phi_{\sss j_1 j_2}^{\sss\rm or}$}

\newcommand\Lzerot{{\cal L}^{\sss t}_0}
\newcommand\LzeroLg{{\cal L}^{\sss\rm loop}_{0,\,g}}
\newcommand\psibt{\bar{\psi}_{\sss t}}
\newcommand\psit{\psi_{\sss t}}

\newcommand\khtt{k_{\sss Htt}}
\newcommand\ghtt{g_{\sss Htt}}
\newcommand\khgg{k_{\sss Hgg}}
\newcommand\ghgg{g_{\sss Hgg}}

\newcommand\katt{k_{\sss Att}}
\newcommand\gatt{g_{\sss Att}}
\newcommand\kagg{k_{\sss Agg}}
\newcommand\gagg{g_{\sss Agg}}
\newcommand\xzero{X_0}
\newcommand\mjj{m_{\sss j_1 j_2}}
\newcommand\Detajj{\Delta\eta_{\sss j_1 j_2}}
\newcommand\Dphijj{\Delta\phi_{\sss j_1 j_2}}

\newcommand\MG{{\sc MadGraph5\_aMC@NLO}}

\newcommand\dontshow[1]{}
\newcommand{\mglong}{{\sc\small MadGraph5\_aMC@NLO}}
\newcommand{\mgshort}{{\sc\small MG5\_aMC}}
\newcommand{\mgfour}{{\sc\small MadGraph4}}
\newcommand{\pwg}{{\sc\small Powheg}}
\newcommand{\gosam}{{\sc\small Gosam}}
\newcommand{\openloops}{{\sc\small OpenLoops}}
\newcommand{\pwgbox}{{\sc\small Powheg Box}}
\newcommand{\pwgboxvtwo}{{\sc\small Powheg Box V2}}
\newcommand{\pwgboxres}{{\sc\small Powheg Box Res}}
\newcommand{\pythia}{{\sc\small Pythia}}
\newcommand{\herwig}{{\sc\small Herwig}}

\newcommand{\mcatnlo}{{{\sc\small MC@NLO}}}

\title{An interface between the  {\sc Powheg Box} and {\sc
    MadGraph5\_aMC@NLO}}

\author[a,b]{Paolo Nason,}
\author[a,b]{Carlo Oleari,}
\author[a,b]{Marco Rocco,}
\author[c]{Marco Zaro}

\emailAdd{paolo.nason@mib.infn.it}
\emailAdd{carlo.oleari@mib.infn.it}
\emailAdd{m.rocco10@campus.unimib.it}
\emailAdd{marco.zaro@mi.infn.it}

\affiliation[a] {Universit\`a di Milano\,-\,Bicocca, Piazza della Scienza 3,
  20126 Milano, Italy}

\affiliation[b] {INFN, Sezione di Milano-Bicocca,
Piazza della Scienza 3, I-20126 Milano, Italy}

\affiliation[c] {INFN, Sezione di Milano \& TIFLab,
Via Celoria 16, I-20133 Milano, Italy}

\abstract{In this paper we present a framework for developing \pwgbox{}
  generators using \MG{} for the computation of the matrix elements.  Within
  this framework, all the flexibility of \MG{} for the generation of matrix
  elements for Standard Model processes and for several of its extensions
  can be exploited, as well as all features of the \pwgbox{} framework,
  including the possibility of multijet merging without a merging scale
  (using the so called MiNLO approach).  As a proof of concept,
  we develop a generator for the production of a spin-0 Higgs-like boson in
  association with up to two jets, with CP-violating couplings.}


\begin{document}

\hfill TIF-UNIMI-2020-20

\maketitle


\section{Introduction}
\label{sec:intro}

Next-to-leading-order~(NLO) calculations for Standard Model~(SM) and,
sometimes, beyond the SM~(BSM) processes, interfaced to parton shower~(PS)
generators, generally dubbed NLO+PS generators, are by now the methods of
choice for the generation of event samples for signal and background
processes at the LHC. This state of the art has been made possible, on the one
side, by the formulations of general methods for computing NLO
corrections~\cite{Frixione:1995ms, Catani:1996vz}, and, on the other, by the
theoretical developement of algorithms for interfacing fixed order
calculations with parton shower generators~\cite{Frixione:2002ik,
  Nason:2004rx, Frixione:2007vw, Jadach:2015mza, Alioli:2012fc,
  Lonnblad:2012ix}. These algorithms were implemented in software packages
for the automatic computation of NLO corrections~\cite{Alwall:2014hca,
  Cascioli:2011va, Cullen:2011ac, Actis:2016mpe, Berger:2008sj}, and for the
automatic implementation of NLO+PS generators~\cite{Alioli:2010xd,
  Alwall:2014hca, Gleisberg:2003xi, Bahr:2008pv, Alioli:2015toa} that
considerably ease the construction of generators for new processes.

\mglong{}, often abbreviated to \mgshort{} in the following, is a framework
where automation has been pushed to the highest level. In fact, a user
without any knowledge of NLO calculations or NLO+PS implementations, can
easily generate samples of parton-level events with NLO+PS accuracy, within
the \mcatnlo{} procedure. These events can be then directly fed into a PS
generator, such as \pythia{} or \herwig{}. The
\mgshort{} framework is not restricted to the case of SM processes.  In fact,
it is possible to employ any user-defined model if this is provided in the
so-called UFO format~\cite{Degrande:2011ua}, for example as generated by {\sc
  FeynRules}~\cite{Christensen:2008py,Alloul:2013bka}.  In particular, in
order to undertake a NLO computation, the model should include the relevant
UV and rational counterterms (both needed for the numerical evaluation of the
one-loop matrix elements), which can be also automatically computed with {\sc
  FeynRules}+{\sc NLOCT}~\cite{Degrande:2014vpa}.
Furthermore, the {\sc FeynRules}+\mgshort{} framework has been recently
extended in order to fully support the supersymmetric case, including the
implementation of different renormalisation
conditions~\cite{Frixione:2019fxg}, and the use of the
so-called diagram removal and diagram subtraction techniques when
intermediate resonances are present.  NLO capabilities for BSM processes have
been proven successful for a number of processes, see
Ref.~\cite{Frixione:2019fxg} and references therein.

The \pwg{} method allows to generate events with positive weights and,
because of this, it has become the method of choice when large samples of
events are needed. In fact, in view of the large amount of computer resources
needed for detector simulation, the experimental collaborations cannot afford
to use the larger samples that are required when negative weights are
present.\footnote{A variant of the \mcatnlo{} method for drastically reducing
  the negative weight fraction has appeared in Ref.~\cite{Frederix:2020trv}.}
The method has been also extended with the introduction of some theoretical
developments of general interest.  One of them deals with the generation of
multijet samples that maintain a certain level of accuracy, even when some of
the jets become unresolved~\cite{Hamilton:2012np, Hamilton:2012rf}. This
approach has also led to the development of NNLO+PS generators,
i.e.~generators where next-to-next-to-leading-order~(NNLO) calculations are
interfaced to parton showers~\cite{Hamilton:2013fea,
  Monni:2019whf}.\footnote{Alternative methods for multijet merging have been
  presented in Refs.~\cite{Frederix:2012ps, Lonnblad:2012ix,
    Alioli:2012fc}. Alternative methods for NNLO+PS accuracy have been
  proposed in Refs.~\cite{Alioli:2013hqa, Hoeche:2014aia}.}  Another development
has been the extension of the \pwg{} method for the inclusion of processes
with decaying coloured resonances, which is capable of handling the interference
of the emitted radiation generated in production and
decay~\cite{Jezo:2015aia}.\footnote{See also Ref.~\cite{Frederix:2016rdc}.}

The \pwgbox{} framework automatises the construction of NLO+PS generators,
once the matrix elements are available. In the early \pwgbox{} processes, the
matrix elements were obtained from the authors of specific calculations. A
considerable leap in the construction of the matrix elements took place when
an interface of the \pwgbox{} to \mgfour{} was set up~\cite{Campbell:2012am},
allowing for the implementation of all tree-level ingredients required by a
given NLO process. After this development, the only missing ingredient for an
NLO calculation in the \pwgbox{} was the virtual contribution. Later,
interfaces to automatic generators of virtual processes were
also developed in Refs.~\cite{Luisoni:2013cuh, Luisoni:2015mpa} for
\gosam{}, and in Ref.~\cite{Jezo:2016ujg} for \openloops.

As of now, an interface to the matrix-element generator that is available
within the \mgshort{} package has not been developed. The main obstacle is
the fact that \mgshort{} is built as a single package that aims at the
production of partonic events, at difference with \mgfour{}, that was
initially conceived for the generation of tree-level matrix elements. An
interface between the matrix-element generator of \mgshort{} and the
\pwgbox{} is also highly desirable since many BSM processes are available
within \mgshort{}. In order to exploit the full capabilities of the
\mgshort{} package, such interface should also build, in addition to the
virtual contribution, all the necessary tree-level matrix elements: the Born,
the colour- and spin-correlated Born, and the real matrix elements.

The purpose of the present work is to present an interface between the
\mgshort{} matrix-element generator and the \pwgbox{}.  The structure of the
interface is such that developments in \mgshort{} and \pwg{} can remain
independent to a large extent. For this reason, our aim is not to construct a
framework that is automatised at the same level as the full \mgshort{}
package itself, but rather to build an \mgshort{} extension that makes the
NLO matrix elements readily available to \pwg{}.  Thus, progresses on the
\pwgbox{} side and on the \mgshort{} side can take place independently, which
is a considerable advantage in view of the way in which theoretical projects
are developed.  Furthermore, this kind of interface allows generalisations to
other NLO+PS frameworks, that may also benefit from it for the implementation
of the matrix elements.

The paper is organised as follows. In Sect.~\ref{sec:interface} we describe
the interface and we give some technical details on how to use it and how to
distribute the generated code.
In Sect.~\ref{sec:Hjj} we consider, as a case study, the production of a
spin-0 boson $X_0$ plus two jets. In particular, we present a few
distributions able to characterise the $X_0$ boson CP properties and we
discuss some features connected to the \pwgbox{} reweighting feature.  We
also show a few distributions obtained with the MiNLO approach.
Finally, in Sect.~\ref{sec:conclusions} we draw our conclusions.


\section{Interface to \mgshort}
\label{sec:interface}
The new interface between \pwg\ and \mgshort\ uses the capability of the
latter to provide tree-level and one-loop matrix elements to be used by the
former. The interface itself is a plugin for \mgshort: as such, it does not
require any modification of the core code and it works with any recent
version of \mgshort.\footnote{Versions~2.6 and onward are fully supported,
  for what concerns QCD corrections.  The extension of the interface to more
  recent releases able to deal with electro-weak corrections (from version
  3)~\cite{Frederix:2018nkq} is left for future work.} It re-organises the
output of \mgshort\ in a format which is suitable for the
\pwgbox~\cite{Alioli:2010xd}, closely
following what is described in Ref.~\cite{Campbell:2012am}. At variance with
what is discussed there, no external providers for the one-loop matrix
elements are needed. Rather, one-loop matrix elements are directly generated
by \mgshort\ thanks to the {\sc MadLoop} module~\cite{Hirschi:2011pa,
  Alwall:2014hca}, which encapsulates several different strategies, such as
integrand reduction~\cite{Ossola:2006us}, Laurent-series
expansion~\cite{Mastrolia:2012bu} and tensor-integral
reduction~\cite{Passarino:1978jh, Davydychev:1991va, Denner:2005nn}, as
implemented in different computer libraries~\cite{Ossola:2007ax,
  Peraro:2014cba, Hirschi:2016mdz, Denner:2016kdg} and improved by an
in-house implementation of the {\sc OpenLoops}
method~\cite{Cascioli:2011va}. Thus, by fully exploiting the capabilities of
{\sc MadLoop}, the evaluation of virtual matrix elements and the assessment
of the numerical stability of the results are granted. Along with the
matrix elements, the relevant helicity routines are also provided, in the
ALOHA format~\cite{deAquino:2011ub}.

\subsection{Technical details}
\label{sec:details}
The interface plugin, dubbed {\sc \plugin}, is publicly available.\footnote{
  \url{https://code.launchpad.net/~mg5amc-pwg-team/mg5amc-pwg/v0}} Its usage
is very simple, as one only needs to copy (or link) the {\tt \pluginpy}
folder inside the {\tt PLUGIN} directory of \mgshort. Please refer to the
{\tt README} file enclosed in the package for conditions of usage and
instructions.

The plugin can be loaded by launching, within the \mgshort{} installation
directory,
\begin{verbatim}
./bin/mg5_aMC --mode=MG5aMC_PWG
\end{verbatim}
in a command shell. In order to generate the code for a specific process at
NLO QCD accuracy, the usual syntax of \mgshort{} should be employed. For
example, in the case of top-pair production, the syntax is the following:
\begin{verbatim}
generate p p > t t~ [QCD]
output pp_ttx
\end{verbatim}
where \verb!pp_ttx! is the name (chosen by the user) of the directory
where the code will be created.  During the execution of the \verb!generate!
command, the {\sc \plugin} plugin checks whether an installation of the
\pwgboxvtwo{} is available on the system and asks for its installation path
(this is needed only once).

When this stage is concluded, the user can \verb!quit! \mgshort{} and finds
the \mgshort{} code for the Born, real and virtual contributions in the
\verb!pp_ttx!  directory, in addition to a few basic \pwgboxvtwo{} files.  In
particular, the \verb!Born.f!, \verb!real.f! and \verb!virtual.f! files are
ready to be used. Also the \verb!init_processes.f! file can be used as it is,
but can be also modified if particular features of the \pwgboxvtwo{} need to
be activated and initialised.

A few comments about the other files are in order:
\begin{itemize}
\item
The \verb!Born_phsp.f! file is just a place holder. It needs to be replaced
by the actual phase-space generator for the process at hand.
Examples of \verb!Born_phsp.f! implementations can be found in the processes
already implemented in the \pwgboxvtwo. In the current setup, a subroutine
\verb!born_suppression! should be also implemented in the \verb!Born_phsp.f!
  file. This function is used at the integration stage to suppress
  divergences when present at the Born level, i.e.~when there are jets and
  photons.

\item
  The call of the \verb!setpara("param_card.dat")! routine in the
  \verb!init_couplings.f! file initialises the parameters listed in the
  \verb!Cards/param_card.dat! file to the corresponding values, according to
  the UFO model~\cite{Degrande:2011ua} used in \mgshort{}.\footnote{It should
    be noted that the \verb!Cards/param_card.dat!  file is not read at
    execution time. Rather, it is parsed at compilation time into a {\sc
      Fortran} include file, which is then compiled together with the
    code. Hence, after any parameter modification within this file, the main
    executable has to be recompiled.}
It is also possible to assign a value to a \mgshort{} parameter at execution
time. An example of this can be found in the \verb!init_couplings.f! file for
the process $X_0jj$, that we discuss in Sect.~\ref{sec:Hjj}.  In this file
we reassign the value of $\cos\alpha$, the CP-mixing parameter that appears
in the Lagrangian of Eq.~(\ref{eq:LHgg}).
This parameter is indicated with \verb!cosa! in the
\verb!Cards/param_card.dat! file, and is initialised to the value specified
in this file, if no further action is taken.
In order to reassign its value at execution time, we change the values of the
internal \mgshort{} variables, \verb!mdl_cosa! and \verb!mp__mdl_cosa!  (for
double and quadruple precision), that encode this parameter.

After any reassignment of the \mgshort{} parameters, the user has to call the
\verb!coup! routine in order to recompute all the dependent variables.

\item


In order to have full consistency between the \mgshort{} amplitudes and what
is computed by the \pwgboxvtwo{}, all the physical parameters used by the
\pwgboxvtwo{} should be set starting from those assigned or computed by
\mgshort{}.  An example of this is the list of the external-particle masses,
\verb!kn_masses!, used by \pwgboxvtwo{} when generating the kinematics of the
event. Using $\ttbar$ production as example, \verb!kn_masses! should be set
to
\begin{verbatim}
(\ 0, 0, mdl_mt, mdl_mt, 0 \)
\end{verbatim}
in \verb!init_couplings.f!  or \verb!Born_phsp.f!, where \verb!mdl_mt! is the
mass of the top quark used inside \mgshort{}, the first two entries are the
masses of the incoming particles, and the last massless particle is the
radiated one, when computing the real contribution.

\item
  The interface also builds a script file, \verb!prepare_run_dir!, that is
  useful to create a directory where the produced code can be executed. For
  example, by typing the command
\begin{verbatim}
./prepare_run_dir test
\end{verbatim}
a directory \verb!test! is created. This directory contains all the relevant
links to the \mgshort{} code and a template of the \verb!powheg.input! file,
required by the \pwgboxvtwo{}.  This last file should then be changed and
modified according to the process at hand.
\end{itemize}
The \pwg{} process generated along these lines can be completed with all
sorts of features that are commonly used in the \pwgboxvtwo.  For example,
one can activate the MiNLO option for processes with associated jets, or use
the damping option to separate the real contributions into two parts, along
the lines of what was suggested in the original \pwg{}
paper~\cite{Nason:2004rx}, and applied for the first time in
Ref.~\cite{Alioli:2008tz}.

\subsection{Distribution of the code}

A process generated with this interface to \mgshort{} cannot be distributed
as a usual \pwgbox{} process, since the searching path of the linked
libraries are written in several files at generation time.

An author can distribute the instructions for \mgshort{}, needed in order to
generate the process, and the actual files, that overwrite the place holders
created by the interface plugin.  In this way, all relevant paths point to
the right directories in the user computer.

Alternatively, the author of the process may provide a script file that
automatically executes all these tasks, helping the installation phase.

\section{A case study: $\boldsymbol{X_0jj}$ production with CP-violating couplings}
\label{sec:Hjj}
For our case study, we considered the production of a spin-0 boson $X_0$ (a
Higgs-like boson) that couples to a massive top quark, produced via gluon
fusion, and accompanied by two jets, in the heavy-top-mass limit. We discuss
a few distributions able to characterise the $X_0$ boson CP properties, and
discuss a few results obtained using the \pwgboxvtwo{} reweighting
feature. We also present a few distributions obtained with the MiNLO method.

\subsection{Theoretical setup}
\label{sec:th}
The theoretical framework of this study is fully inherited from what was
done in Ref.~\cite{Demartin:2014fia}, where the process was studied at NLO in
QCD. In particular, in the heavy-top-mass limit, the CP structure of the
$X_0$-top interaction characterises the effective $gg X_0$ vertex. The
starting point is the effective Lagrangian 
\begin{equation}
  \label{eq:LHttbar}
  \Lzerot = -\psibt \( \khtt\,\ghtt\,\cos\alpha +
  i\,\katt\,\gatt\,\sin\alpha\,\gamma_5 \) \psit\,\xzero \,,
\end{equation}
where $\xzero$ is the spin-0 boson, $\psit$ the top-quark spinor, $\alpha$
the CP-mixing angle parameter ($0 \le \alpha \le \pi$), $\khtt$ and $\katt$
the real  coupling parameters and
\begin{equation}
  \ghtt = \gatt = \frac{m_{\sss t}}{v} = \frac{y_{\sss t}}{\sqrt{2}} 
\end{equation}
the Yukawa couplings, with $v$ the vacuum expectation value.

The CP-even case, that will be labeled $0^+$, corresponds to the assignment
$\cos\alpha = 1$, namely to the SM scenario, while the CP-odd case, labeled
$0^-$, to $\cos\alpha = 0$. A CP-mixed case, $0^\pm$, where the spin-0 boson
receives contributions from both a scalar and a pseudoscalar state, is also
taken into account by setting $\cos\alpha = 1/\sqrt{2}$.

For our purposes, it will suffice to notice that the Higgs interaction with
the gluons originates as an effective coupling induced by a top-quark
loop. The relevant effective Lagrangian, in the {\it Higgs Characterisation}
framework~\cite{Artoisenet:2013puc}, reads
\begin{equation}
  \label{eq:LHgg}
  \LzeroLg = -\frac{1}{4} \(
  \khgg\,\ghgg\,\cos\alpha\,\,G^a_{\mu\nu}\,G^{a,\mu\nu} +
  \kagg\,\gagg\,\sin\alpha\,\,
  \epsilon^{\mu\nu\rho\sigma}\,G^a_{\mu\nu}\,G^a_{\rho\sigma} \) \xzero
  \,,
\end{equation}
where $G^a_{\mu\nu}$ is the gluon field strength and
\begin{equation}
  \khgg = -\frac{\as}{3\pi v} \,, \qquad\qquad \kagg = \frac{\as}{2\pi v} \,.
\end{equation}
%
The theoretical setup is made available online in the {\sc
  FeynRules}~\cite{Alloul:2013bka} repository as a UFO model named {\tt
  HC\_NLO\_X0}~\cite{Maltoni:2013sma, Demartin:2014fia, Demartin:2015uha,
  Demartin:2016axk}, which is in fact the one used for our case study.

\subsection{Generation of the code} 
In order to generate the code, we have first to install the UFO model
\verb!HC_NLO_X0_UFO.zip! under the \verb!models! directory of the \mgshort{}
version being used.  We have then followed the procedure described in
Sect.~\ref{sec:details} for the generation of the code, and given the
following commands to \mgshort{}:
\begin{verbatim}
import model HC_NLO_X0_UFO-heft
generate p p > x0 j j / t [QCD]
install ninja
install collier
output X0jj
\end{verbatim}
where we have also inserted the command lines to install {\sc
  ninja}~\cite{Peraro:2014cba} and {\sc collier}~\cite{Denner:2016kdg}, that
are optional and need to be installed just once.

We have then overwritten the \verb!Born_phsp.f! file generated by the
interface with the \verb!Born_phsp.f! from the $Hjj$ \pwgboxvtwo{} process,
taking care of assigning to the \pwg{} variables \verb!hmass! and
\verb!hwidth! (the mass and width of the Higgs-like boson) the \mgshort{}
values, \verb!mdl_mx0! and \verb!mdl_wx0! respectively.

In order to ease the installation procedure, we provide a tarball file that
needs to be inflated in the installation directory. This file contains all
the modified files that replace the place holders.

\subsection{Simulation parameters}
We have performed a simulation for the LHC, running at a centre-of-mass
energy of $\sqrt{S}=13$~TeV. The mass of the spin-0 boson
$\xzero$ has been set equal to 125~GeV. We have chosen the NNPDF2.3~(NLO)
set~\cite{Ball:2012cx} for the parton distribution functions, within the
LHAPDF interface~\cite{Whalley:2005nh, Buckley:2014ana}.

The differential cross section for $X_0 jj$ production is already divergent
at the Born level, unless a minimum set of generation cuts is imposed on the
transverse momentum of the final-state jets and on their invariant
mass. Alternatively, the divergences can be avoided if the code is executed
with the MiNLO option activated.  We have generated the kinematics of the
underlying Born configurations imposing the following minimum set of cuts
\begin{equation}
 p_{\sss\rm T}^{\sss j_k} >10~{\rm GeV}\,, \quad k = 1,2 \,, \qquad\qquad
 \mjj > 10~{\rm GeV}\,.
\end{equation}
In the phenomenological study we perform in Sect.~\ref{sec:pheno}, we apply
more stringent cuts, and we have checked that the results we present are
insensitive to the generation cuts.

In order to integrate the divergent underlying Born cross section, the
\pwgboxvtwo{} can further apply a suppression factor at the integrand
level. We stress that the final kinematic distributions are independent of
this factor.\footnote{We have set {\tt bornsuppfact} to 30~GeV in our
  simulation.}

\subsection{Phenomenology}
\label{sec:pheno}
In this section we present results produced by the \pwgboxvtwo{} at the Les
Houches Event~(LHE) level, i.e.~after the emission of the first radiation,
accurate at NLO for large transverse momentum, and with leading-logarithmic
accuracy at small $\pt$, due to the presence of the \pwg{} Sudakov form
factor.  The results are computed on samples of 3.2~M events.

The renormalisation and factorisation scales are set to
\begin{equation}
  \mu_{\sss\rm R} = \mu_{\sss\rm F} = \frac{H_{\sss\rm T}}{2} \,,
\end{equation}
where $H_{\sss\rm T}$ is the sum of the transverse masses of the particles in
the final state.

Jets are reconstructed employing the anti-$\kt$
algorithm~\cite{Cacciari:2008gp} via the {\sc FastJet}
implementation~\cite{Cacciari:2011ma}, with distance parameter $R=0.4$, and
the two leading jets are required to have transverse momentum and
pseudorapidity such that
\begin{equation}
  p_{\sss\rm T}^{\sss j_k} >30~{\rm GeV}\,, \qquad   |\eta_{\sss j_k}| <
  4.5 \,,\qquad k = 1,2 \,.
\end{equation}
Events that do not pass this minimum set of acceptance cuts are discarded.

\begin{figure}[htb!]
  \begin{center}
    \includegraphics[width=0.7\textwidth]{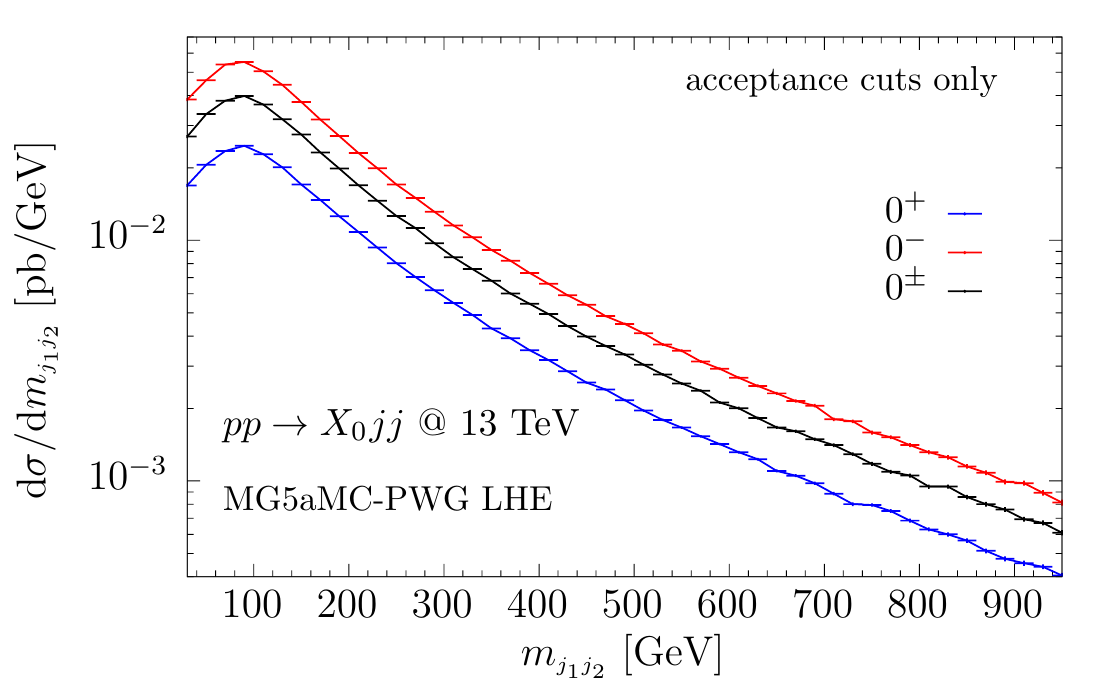}
  \end{center}
  \caption{Differential cross section as a function of the invariant-mass
    distribution of the two leading jets in $pp\to\xzero\hspace{0.5pt}jj$ for
    the three CP scenarios. The blue curve corresponds to the CP-even
    scenario with $\cos\alpha = 1$, the red curve to the CP-odd scenario with
    $\cos\alpha = 0$ and the black curve to the mixture of the $0^+$ and
    $0^-$ scenarios with $\cos\alpha = 1/\sqrt{2}$.}
  \label{fig:mj1j2}
\end{figure}
In Fig.~\ref{fig:mj1j2} we plot the differential cross section for $X_0jj$
production as a function of the invariant mass of the two leading jets,
$m_{j_1 j_2}$, for three different CP scenarios: CP even ($0^+$), CP odd
($0^-$) and a mixture of the two ($0^\pm$).  The shapes of the three spectra
are very similar among each other.  Since a cut on the invariant mass of the
dijet system enhances the discriminating power among different CP
scenarios~\cite{Hagiwara:2009wt}, the fact that the three spectra have
similar shapes implies that the cut acts in a similar way on each of them.
Typically a cut on $m_{j_1 j_2}$ enhances the contributions coming from the
exchange of a gluon in the $t$ channel, and these contributions are more
sensitive to the CP properties of the $X_0$ boson.

\begin{figure}[htb!]
  \begin{center}
    \includegraphics[width=0.495\textwidth]{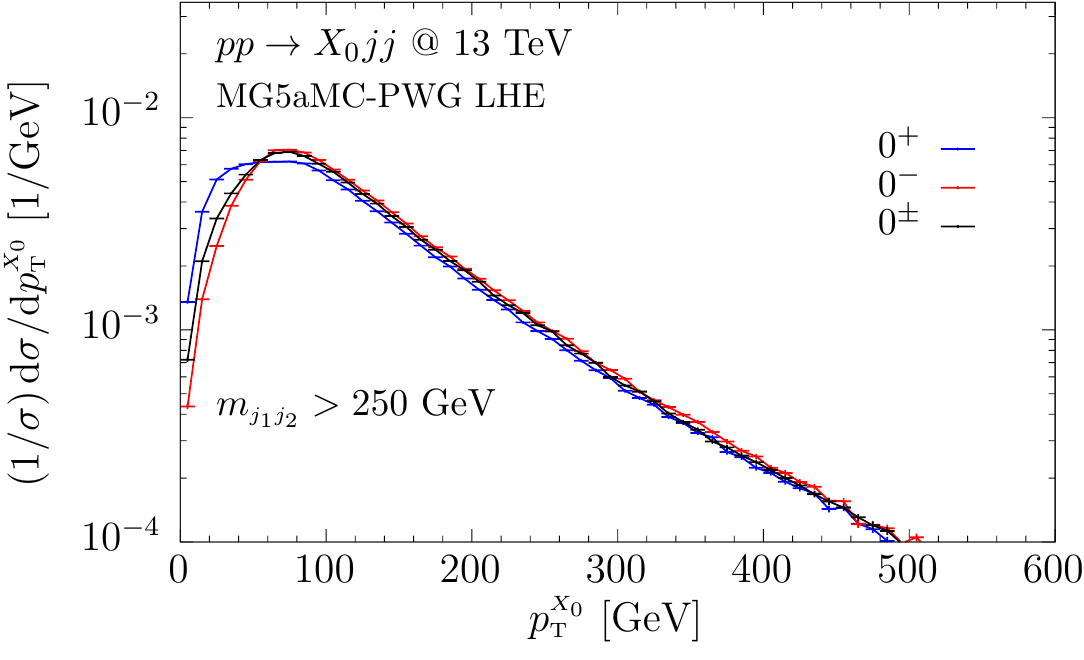}
    \includegraphics[width=0.495\textwidth]{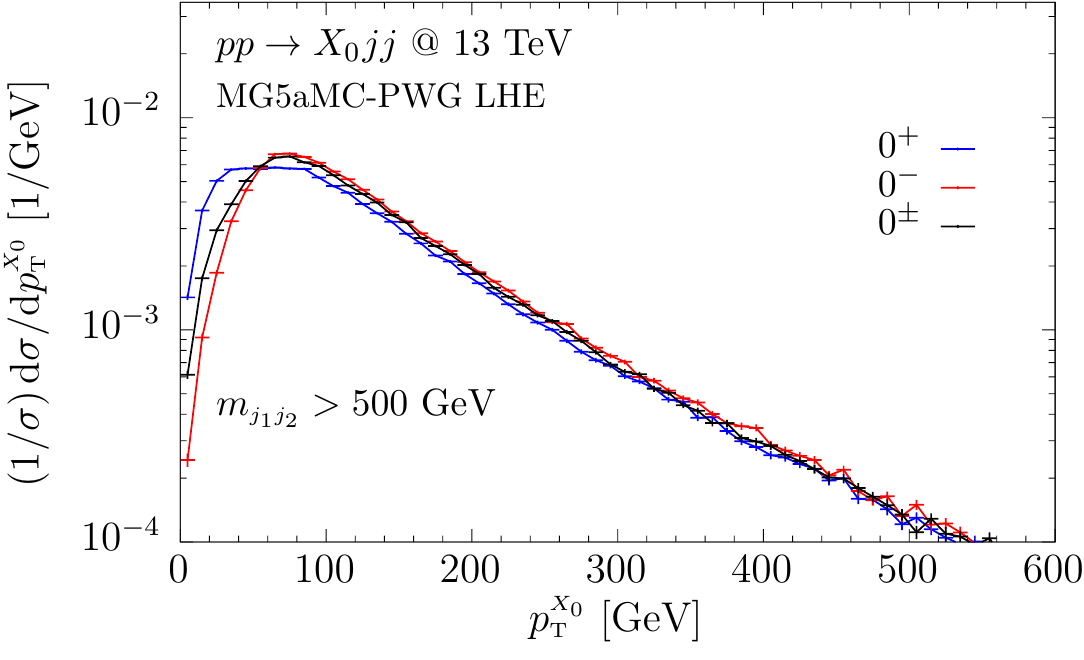}
  \end{center}
  \caption{Normalised differential cross section as a function of the
    transverse momentum of the spin-0 boson $\xzero$, for the three CP
    scenarios. On the left pane, a cut of 250~GeV is imposed on the dijet
    mass, while on the right pane a cut of 500~GeV is applied. The colour code
    is the same as in Fig.~\ref{fig:mj1j2}.}
  \label{fig:Hpt}
\end{figure}

\begin{figure}[htb!]
  \begin{center}
    \includegraphics[width=0.495\textwidth]{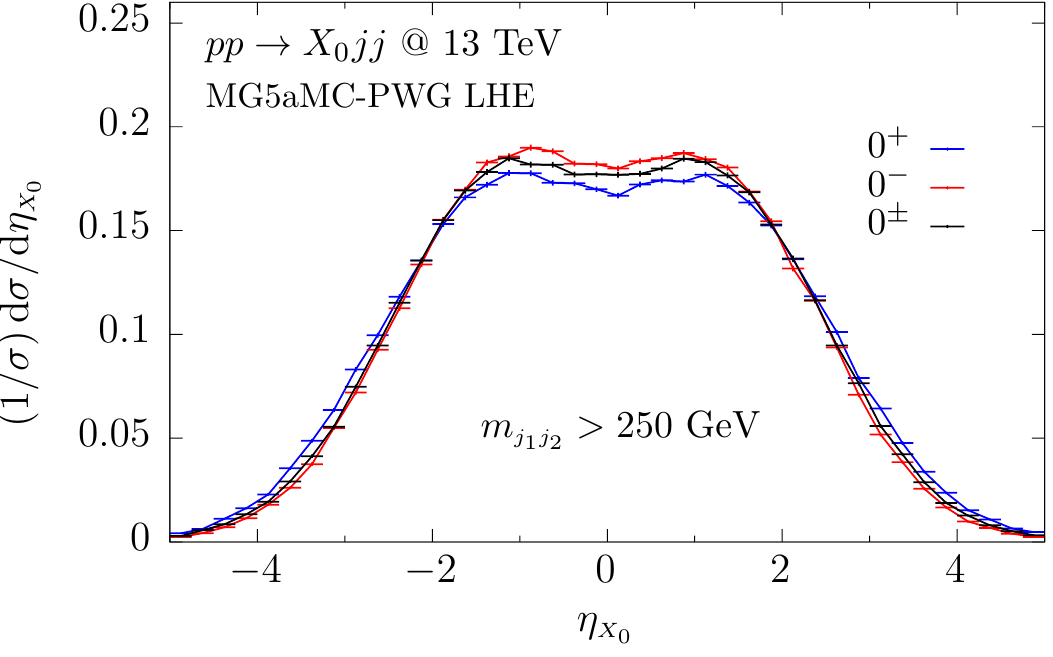}
    \includegraphics[width=0.495\textwidth]{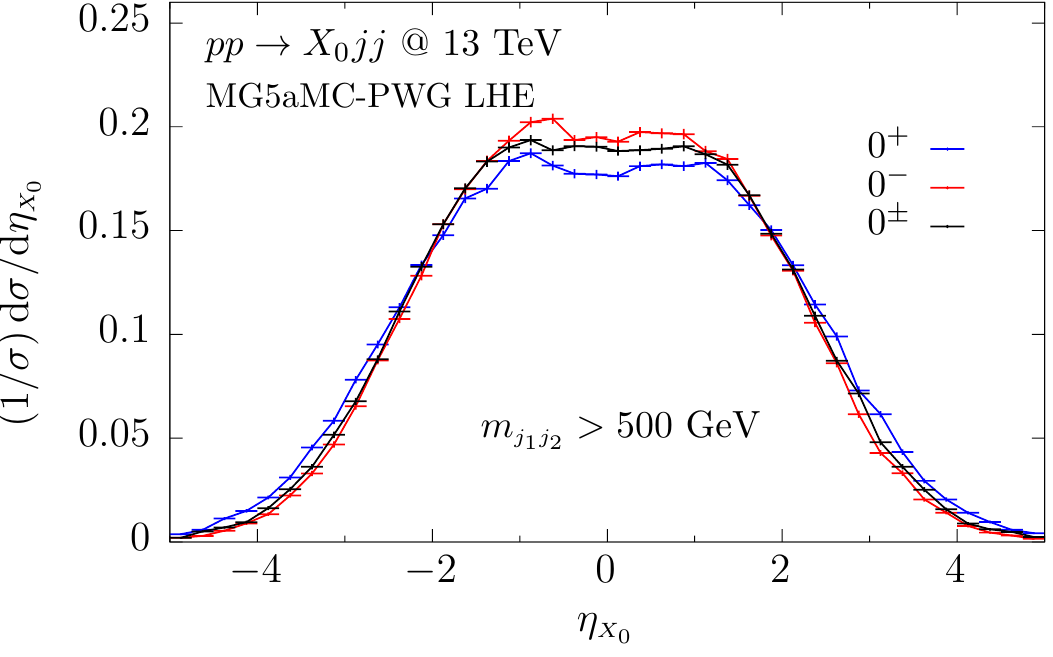}
  \end{center}
  \caption{Normalised differential cross section as a function of the
    pseudorapidity of the spin-0 boson $\xzero$, for the three CP
    scenarios. On the left pane, a cut of 250~GeV is imposed on the dijet
    mass, while on the right pane a cut of 500~GeV is applied. The colour code
    is the same as in Fig.~\ref{fig:mj1j2}.}
   \label{fig:Heta}
\end{figure}

\begin{figure}[htb!]
  \begin{center}
    \includegraphics[width=0.495\textwidth]{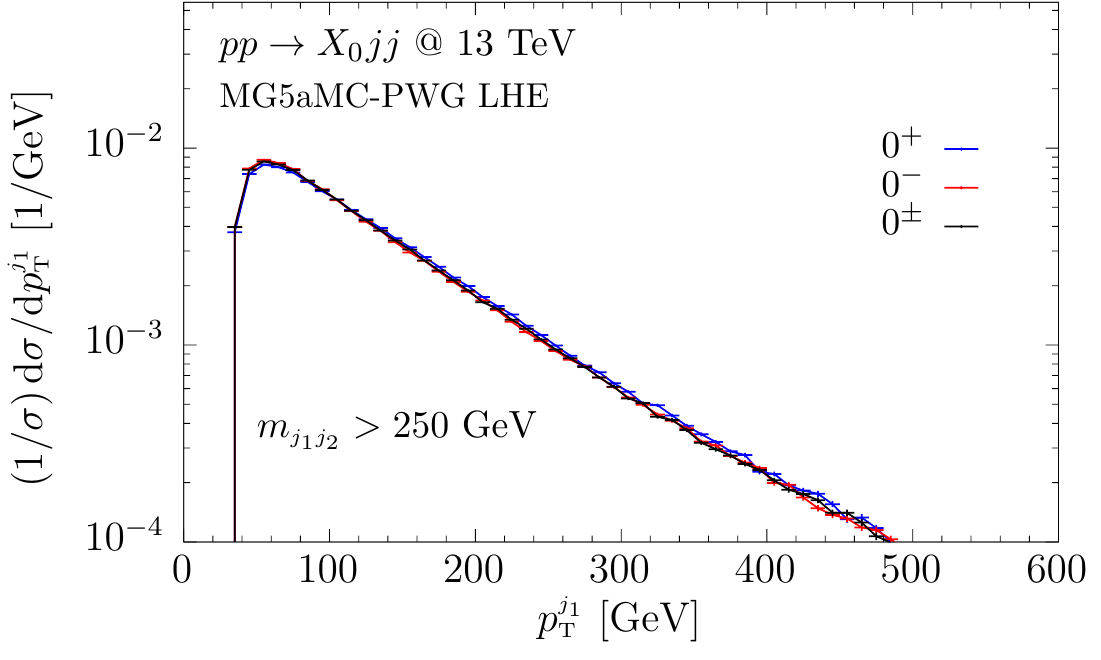}
    \includegraphics[width=0.495\textwidth]{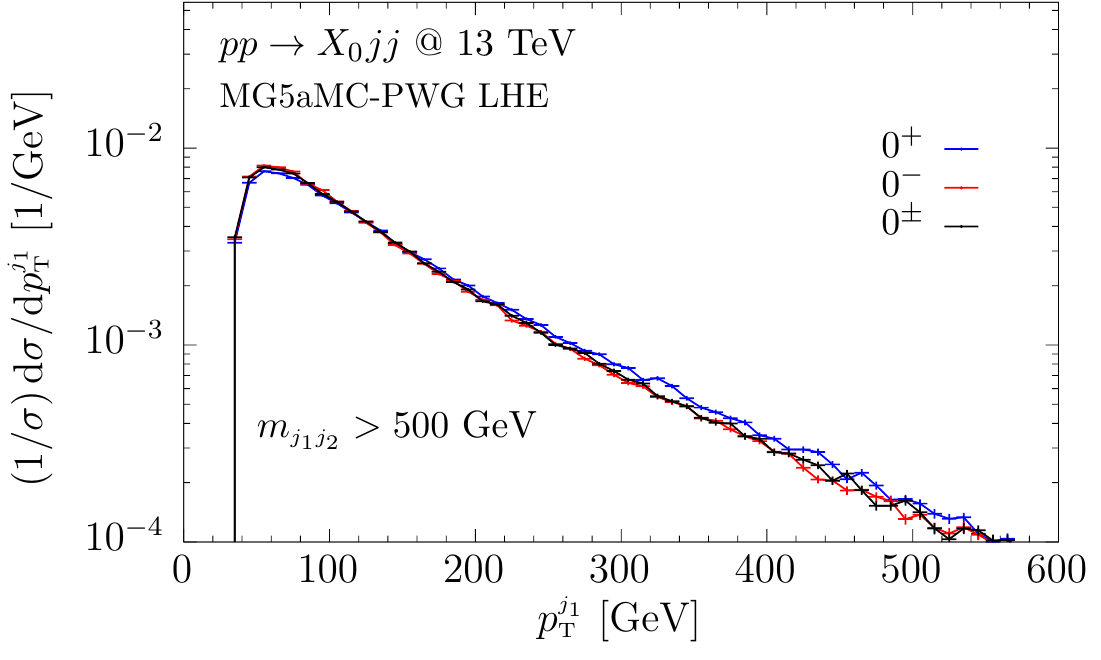}
  \end{center}
  \caption{Normalised differential cross section as a function of the
    transverse momentum of the leading jet, for the three CP scenarios. On
    the left pane, a cut of 250~GeV is imposed on the dijet mass, while on
    the right pane a cut of 500~GeV is applied. The colour code is the same as
    in Fig.~\ref{fig:mj1j2}.}
  \label{fig:j1pt}
\end{figure}

\begin{figure}[htb!]
  \begin{center}
    \includegraphics[width=0.495\textwidth]{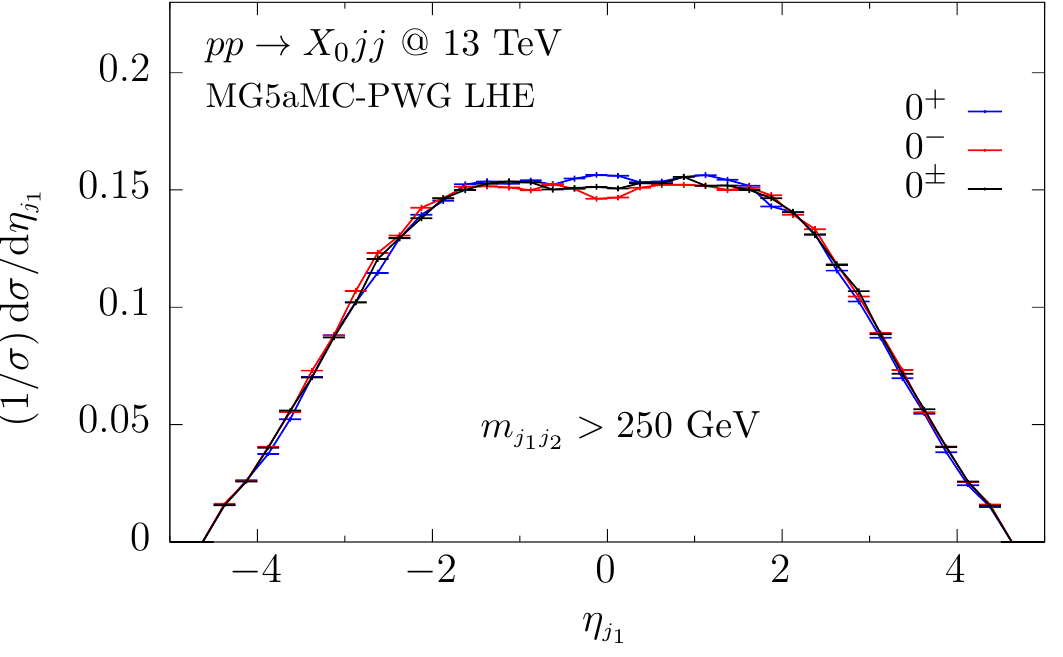}
    \includegraphics[width=0.495\textwidth]{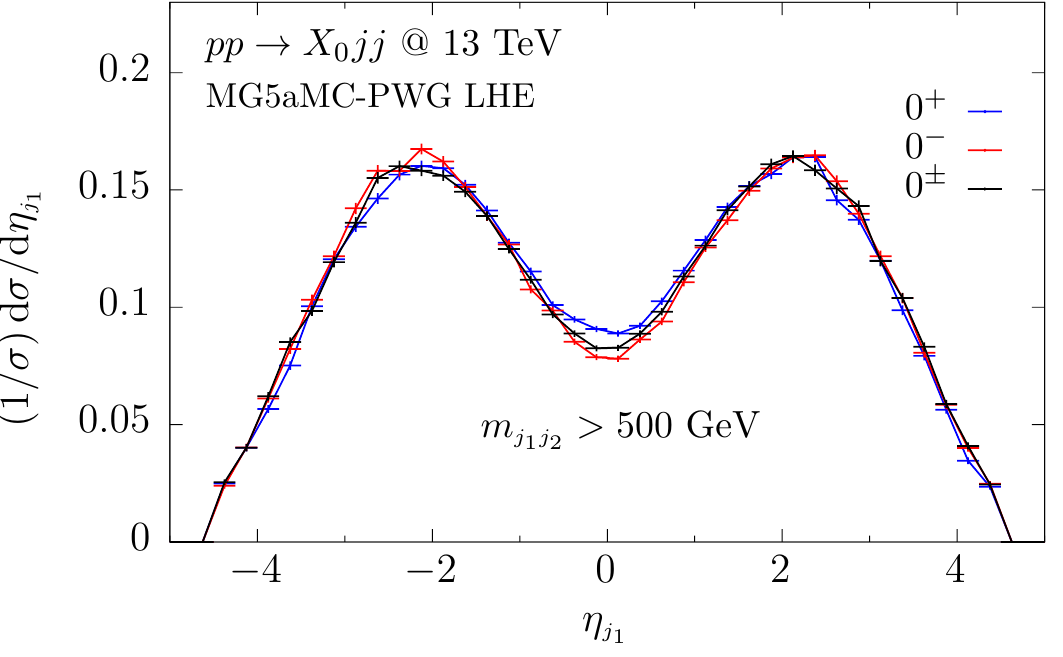}
  \end{center}
 \caption{Normalised differential cross section as a function of the
   pseudorapidity of the leading jet, for the three CP scenarios. On the left
   pane, a cut of 250~GeV is imposed on the dijet mass, while on the right
   pane a cut of 500~GeV is applied. The colour code is the same as in
   Fig.~\ref{fig:mj1j2}.}
  \label{fig:j1eta}
\end{figure}

In the following plots we impose an additional cut on the dijet mass. In
particular, we consider the two cases where
\begin{equation}
  \mjj > 250 \,\,{\rm GeV} \qquad {\rm and} \qquad \mjj>500 \,\, {\rm GeV} \,.
\end{equation}
In addition, since we are interested in shape comparisons among different CP
scenarios, we normalise each curve to one.

In Figs.~\ref{fig:Hpt} and~\ref{fig:Heta} we plot the transverse momentum and
pseudorapidity of the $X_0$ boson, and in Figs.~\ref{fig:j1pt}
and~\ref{fig:j1eta} we show the transverse momentum and pseudorapidity of the
leading jet.  The increase of the cut on the dijet mass hardens the $\pt$
spectrum of the $X_0$ boson and the leading jet $j_1$.  Moreover, there are
only mild differences among the three CP scenarios in the $X_0$ distributions
at low transverse momentum and in the central pseudorapidity region, with a
modest enhancement when the dijet-mass cut increases. No substantial
differences are present in $p_{\sss\rm T}^{\sss j_1}$ and $\eta_{\sss j_1}$,
also in agreement with what is found in
Ref.~\cite{Demartin:2014fia}.\footnote{A possible concern is to what extent
  the effective-field-theory~(EFT) Lagrangian of Eq.~(\ref{eq:LHgg}) produces
  sound results in the high-energy regimes, since it describes the full
  theory in the heavy-top-quark limit. From the exact calculation of
  Ref.~\cite{DelDuca:2001fn}, it is known that the EFT closely reproduces the
  $\mjj$ spectrum even in the very high invariant-mass region. However, the
  EFT approximation breaks down when the transverse momenta of the jets are
  larger than the top mass~\cite{DelDuca:2001eu}, overestimating the exact
  prediction when $p_{\sss \rm T}^{\sss j_1}$ is larger than the top
  mass. Since the region of interest for discriminating the CP properties is
  at low transverse momentum, we can trust the results obtained within the
  EFT approach.}

\begin{figure}[htb!]
  \begin{center}
    \includegraphics[width=0.495\textwidth]{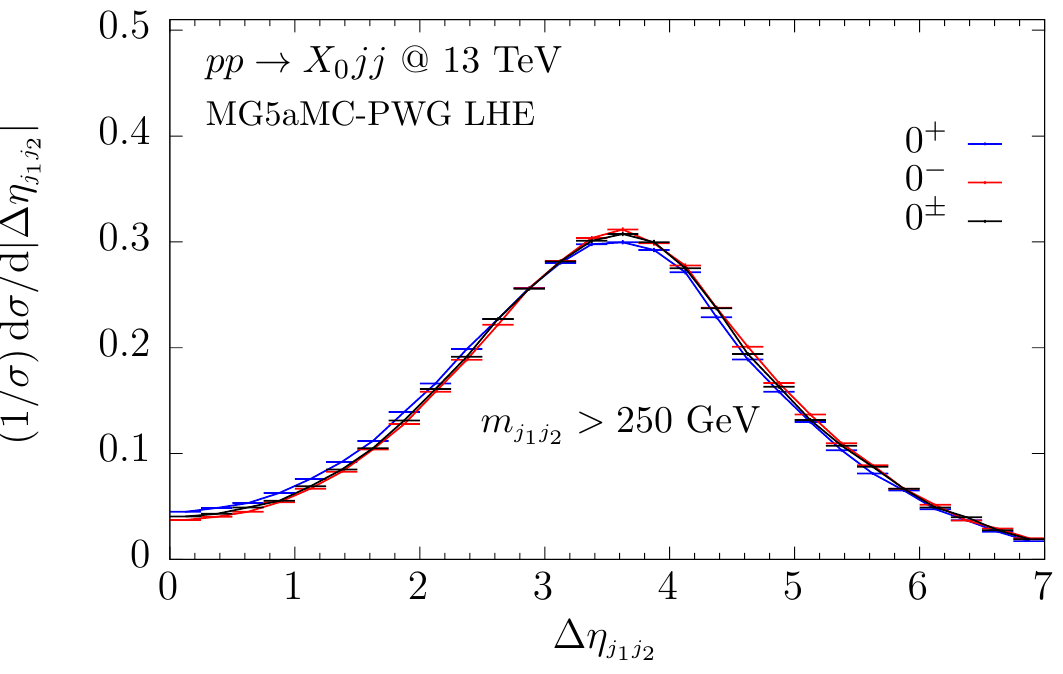}
    \includegraphics[width=0.495\textwidth]{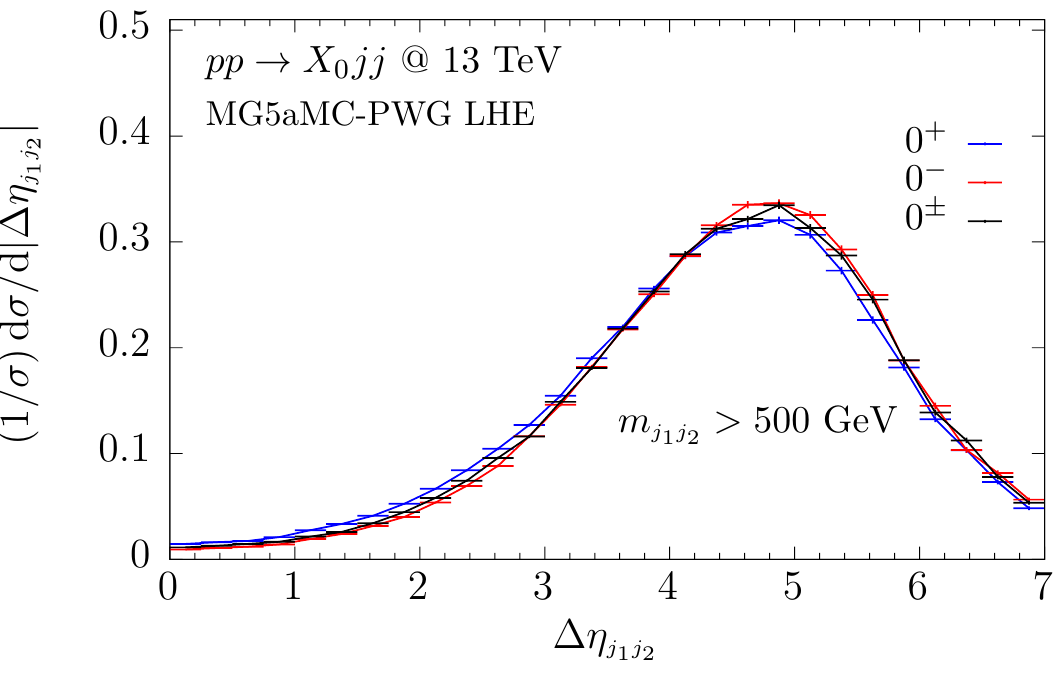}
  \end{center}
 \caption{Normalised differential cross section as a function of the
   pseudorapidity separation of the two leading jets (see
   eq.~(\ref{eq:deltaetaj1j2})), for the three CP scenarios. On the left pane,
   a cut of 250~GeV is imposed on the dijet mass, while on the right pane a
   cut of 500~GeV is applied. The colour code is the same as in
   Fig.~\ref{fig:mj1j2}.}
  \label{fig:j1j2deta}
\end{figure}

\begin{figure}[htb!]
  \begin{center}
    \includegraphics[width=0.495\textwidth]{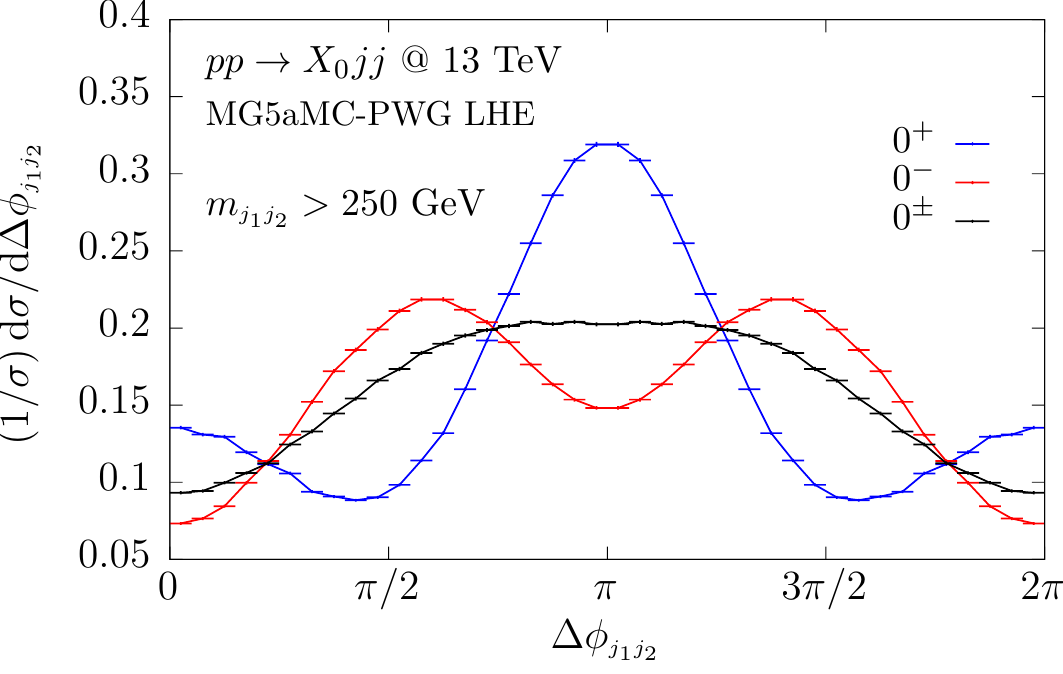}
    \includegraphics[width=0.495\textwidth]{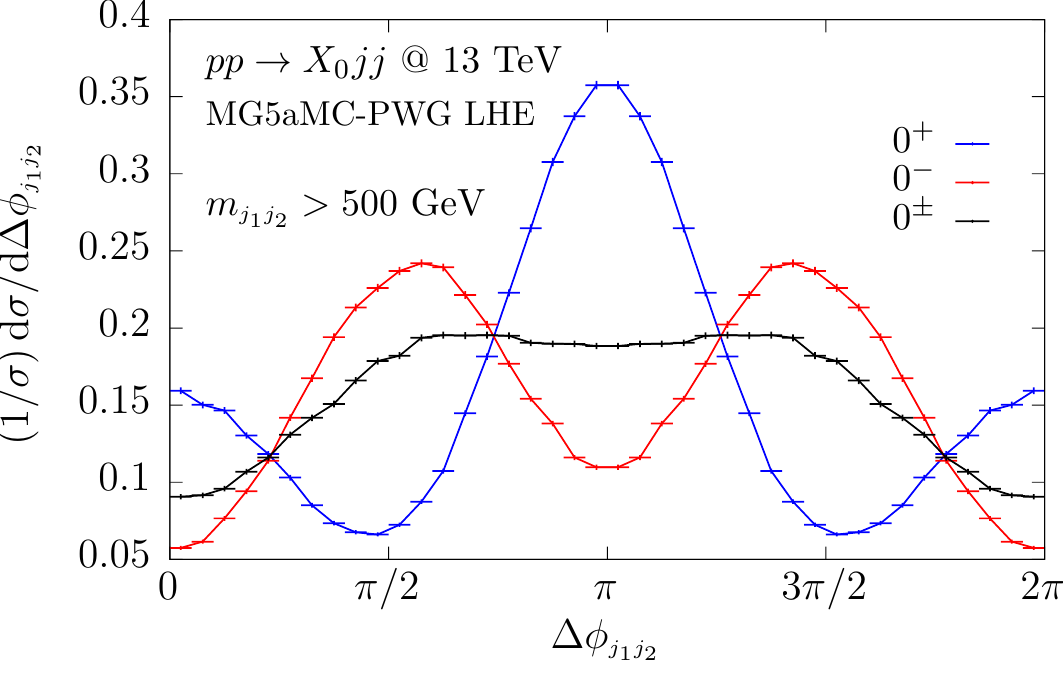}
  \end{center}
  \caption{Normalised differential cross section as a function of the
    azimuthal separation of the two leading jets (see
   eq.~(\ref{eq:deltaphij1j2})), for the three CP
    scenarios. On the left pane, a cut of 250~GeV is imposed on the dijet
    mass, while on the right pane a cut of 500~GeV is applied. The colour code
    is the same as in Fig.~\ref{fig:mj1j2}.}
  \label{fig:j1j2dphi}
\end{figure}

The most sensitive observables to the CP coupling of the $X_0$ boson to the
top quark in gluon fusion are dijet-correlation variables~\cite{Plehn:2001nj,
  Klamke:2007cu, Hagiwara:2009wt, Andersen:2010zx, Campanario:2010mi,
  Englert:2012ct, Englert:2012xt, Dolan:2014upa}.  As displayed in
Fig.~\ref{fig:j1j2deta}, no significant differences are seen in the
differential cross sections as a function of the pseudorapidity separation of
the two leading jets
\begin{equation}
  \label{eq:deltaetaj1j2}
  \Detajj = \left|\eta_{\sss j_1} - \eta_{\sss j_2}\right|.
\end{equation}
Instead, when the differential cross sections are expressed as a function of
the azimuthal-angle separation, the CP nature of the coupling is more
evident~\cite{Hagiwara:2009wt}.  In fact, the shape of the differential cross
sections as a function of $\Dphijj$ are very different, as shown in
Fig.~\ref{fig:j1j2dphi}, where we have defined (modulo $2\pi$)
\begin{equation}
  \label{eq:deltaphij1j2}
  \Dphijj = \left|\phi_{\sss j_1} - \phi_{\sss j_2}\right| ,
\end{equation}
where the azimuth of a jet is computed as
\begin{equation}
  \phi_{\sss j_k} = \arg \( {\bf p}^{j_k} \cdot \hat{y} + i\, {\bf
        p}^{j_k} \cdot \hat{x} \), \quad k=1,2\,,
\end{equation}
with ${\bf p}^{j_k}$ the tri-momentum of the jet $k$ and $\hat{x}\,(\hat{y})$ the
unit vector along the $x\,(y$)-axis direction.

\begin{figure}[htb!]
  \begin{center}
    \includegraphics[width=0.495\textwidth]{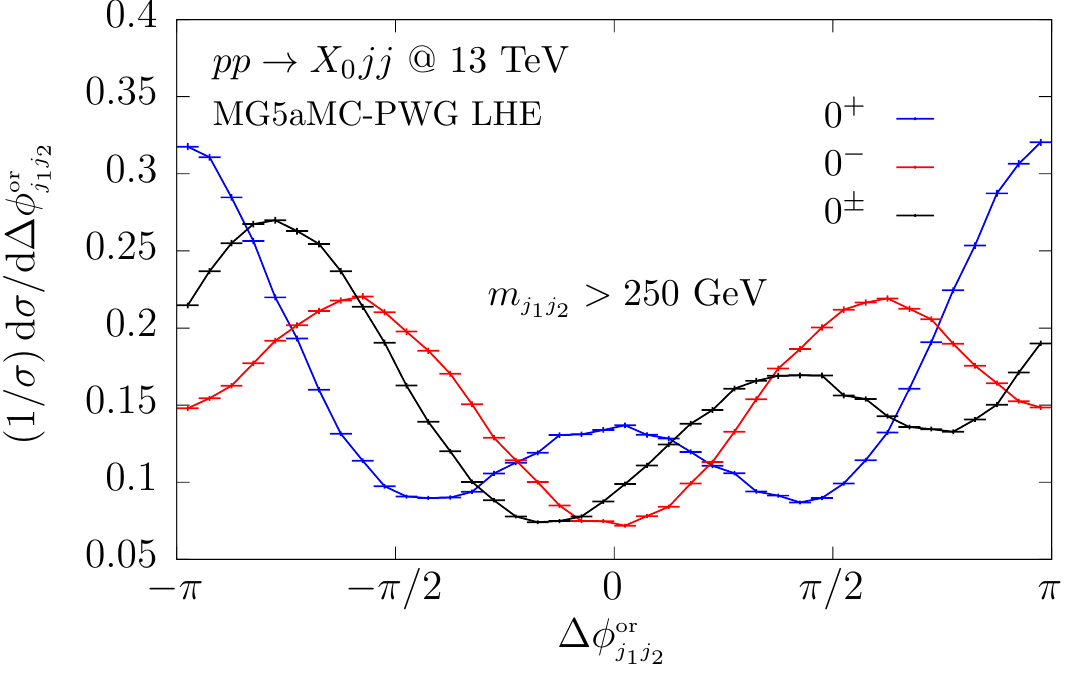}
    \includegraphics[width=0.495\textwidth]{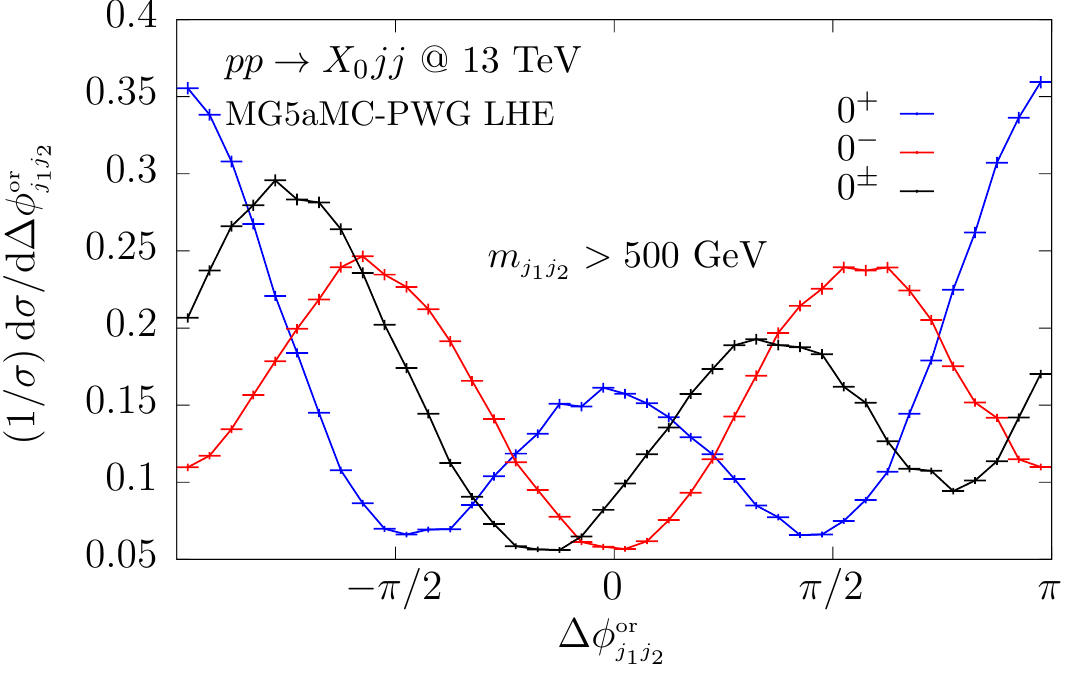}
  \end{center}
 \caption{Normalised differential cross section as a function of the oriented
   azimuthal separation of the two leading jets, defined in
   Eq.~(\ref{eq:deltaphij1j2or}), for the three CP scenarios. On the left
   pane, a cut of 250~GeV is imposed on the dijet mass, while on the right
   pane a cut of 500~GeV is applied. The colour code is the same as in
   Fig.~\ref{fig:mj1j2}.}
 \label{fig:deltaphij1j2or}
\end{figure}

As pointed out in Refs.~\cite{Klamke:2007cu,Hankele:2006ma},
a more CP-sensitive observable (especially for the maximal mixing scenario of
$\cos\alpha=1/\sqrt{2}$ considered here) is the oriented azimuthal separation
of the two hardest jets. This variable contains information not only on the
azimuthal separation of the two jets but also on the sign of the azimuthal
angle.  We have adopted the definition of this variable of
Ref.~\cite{Gritsan:2020pib}, namely
\begin{equation}
  \label{eq:deltaphij1j2or}
  \Delta\phi_{\sss j_1 j_2}^{\sss\rm or} \equiv
  \frac{\(\hat{\bf p}_{\sss \rm T}^{\sss j_1} \times \hat{\bf p}_{\sss \rm T}^{\sss j_2}\)
    \cdot \hat{z}}
   {\left|\(\hat{\bf p}_{\sss \rm T}^{\sss j_1} \times \hat{\bf p}_{\sss \rm T}^{\sss j_2}\)
     \cdot \hat{z} \right|}
   \, \frac{\({\bf p}^{\sss j_1} - {\bf p}^{\sss j_2}\)\cdot \hat{z}}
      {\left| \({\bf p}^{\sss j_1} - {\bf p}^{\sss j_2}\)\cdot \hat{z} \right|}
      \,\arccos\!{\(\hat{\bf p}_{\sss \rm T}^{\sss j_1} \cdot \hat{\bf p}_{\sss \rm T}^{\sss j_2}\)} \,,
\end{equation}
where $\hat{\bf p}_{\sss \rm T}^{\sss j_k}$ is the jet transverse momentum,
normalised to one, and $\hat{z}$ is the unit vector along the $z$-axis
direction.

The differential cross sections for the three different CP scenarios
considered in this paper, as a function of \xDphiorjj, are shown in
Fig.~\ref{fig:deltaphij1j2or}, and their shape is visibly different.

\begin{figure}[htb!]
  \begin{center}
    \includegraphics[width=0.7\textwidth]{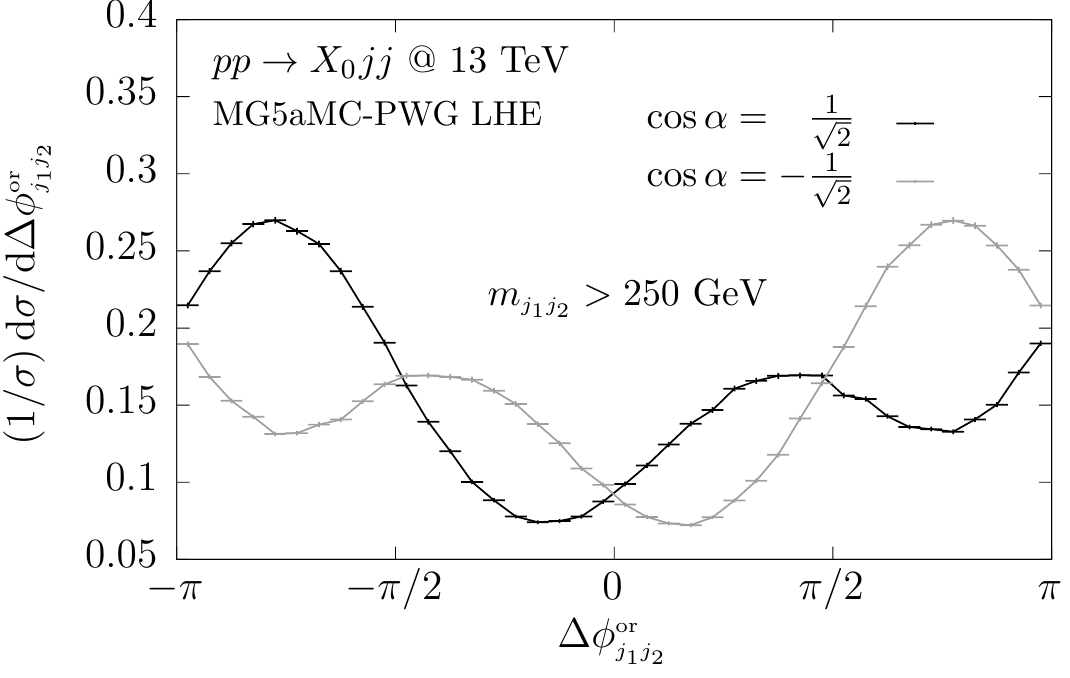}
  \end{center}
\caption{Normalised differential cross section as a function of the oriented
  azimuthal separation of the two leading jets, defined in
  Eq.~(\ref{eq:deltaphij1j2or}), for the two mixed CP scenarios with
  $\cos\alpha=1/\sqrt{2}$ (black curve) and $\cos\alpha=-1/\sqrt{2}$ (grey
  curve). A cut of 250~GeV is imposed on the dijet mass.}
  \label{fig:j1j2dphior-0pm}
\end{figure}

In particular, the oriented azimuthal separation can also distinguish between
the two scenarios with $\cos\alpha=1/\sqrt{2}$ and $\cos\alpha=-1/\sqrt{2}$,
as illustrated in Fig.~\ref{fig:j1j2dphior-0pm}, while \xDphijj cannot
distinguish between them.

\subsection{Reweighting}

\begin{figure}[htb!]
  \begin{center}
    \includegraphics[width=0.495\textwidth]{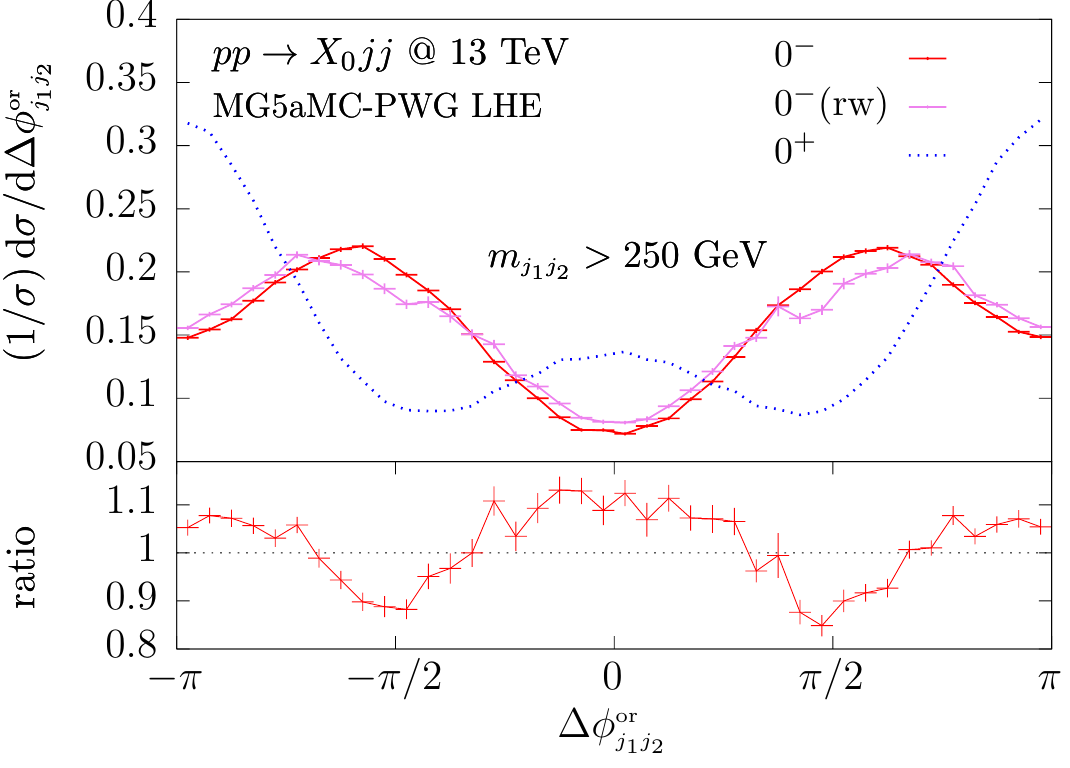}
    \includegraphics[width=0.495\textwidth]{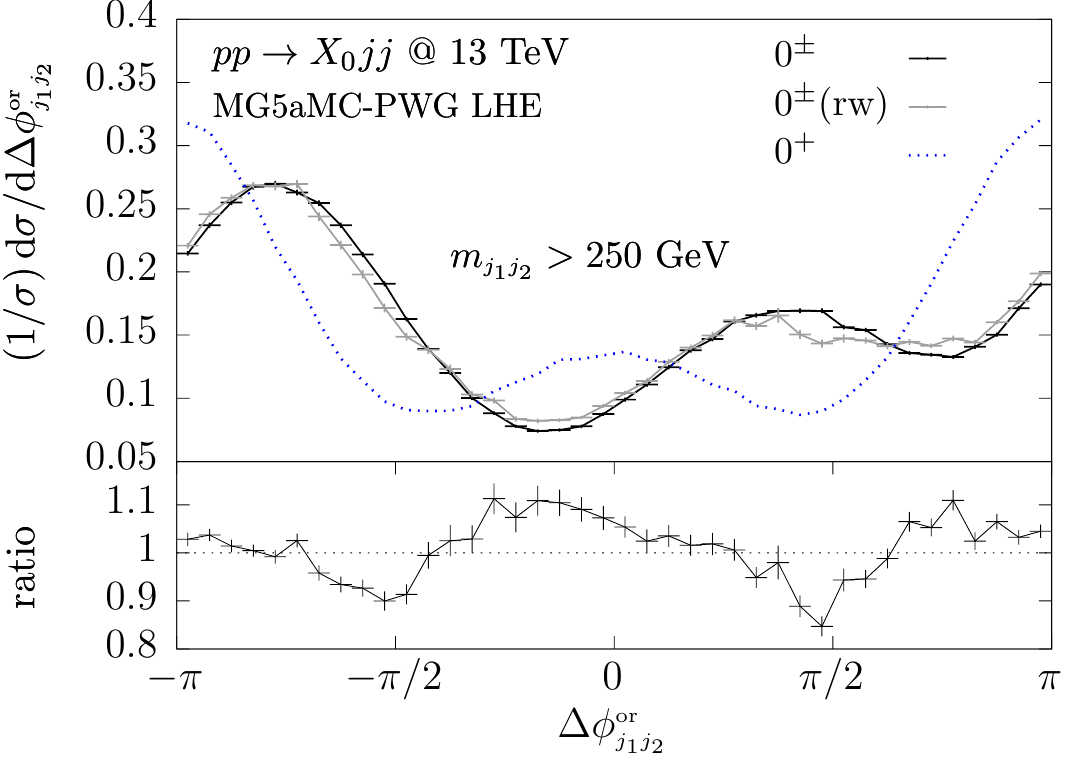}
  \end{center}
  \caption{Normalised differential cross section as a function of the
    oriented azimuthal separation of the two leading jets, defined in
    Eq.~(\ref{eq:deltaphij1j2or}), with a cut of 250~GeV imposed on the dijet
    mass. On the left pane, the pseudoscalar original distribution in red,
    the pseudoscalar as obtained by reweighting~(rw) in pink, and the scalar
    one in dotted blue.  On the right pane, the CP mixed original
    distribution in black, the mixed as obtained by reweighting~(rw) in gray,
    and the scalar one in dotted blue. The ratios between the distributions
    obtained by reweighting and the original ones are also shown.}
   \label{fig:rwgt_from_scalar}
\end{figure}

\begin{figure}[htb!]
  \begin{center}
    \includegraphics[width=0.495\textwidth]{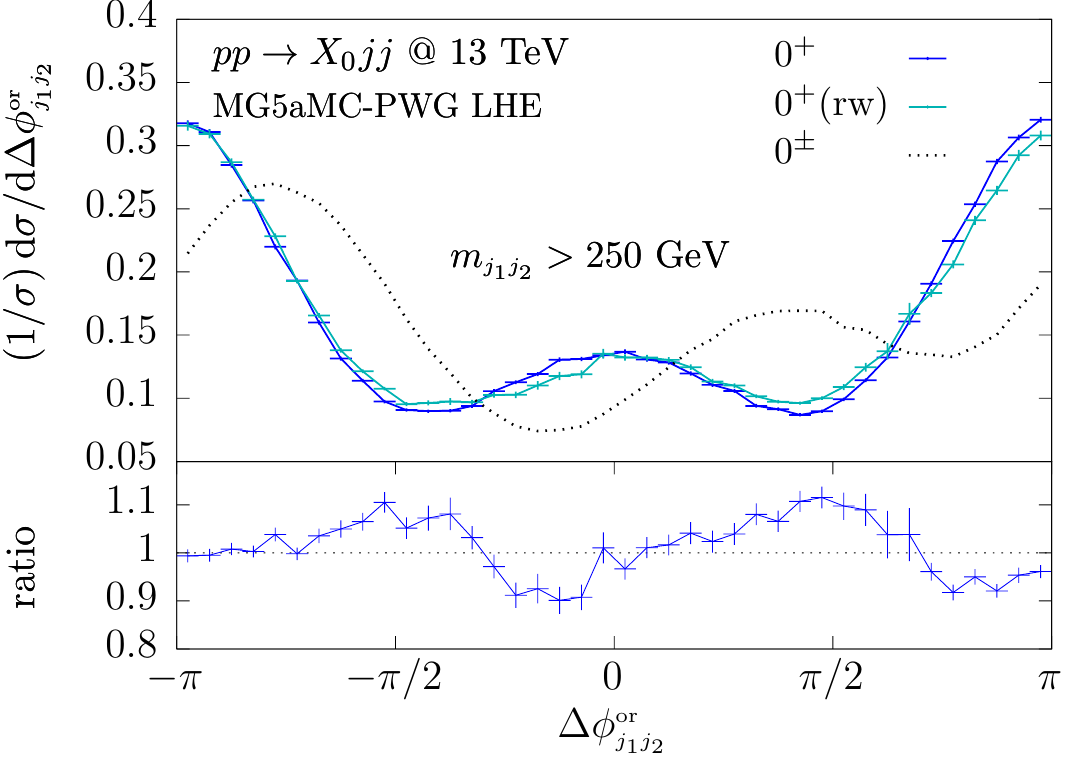}
    \includegraphics[width=0.495\textwidth]{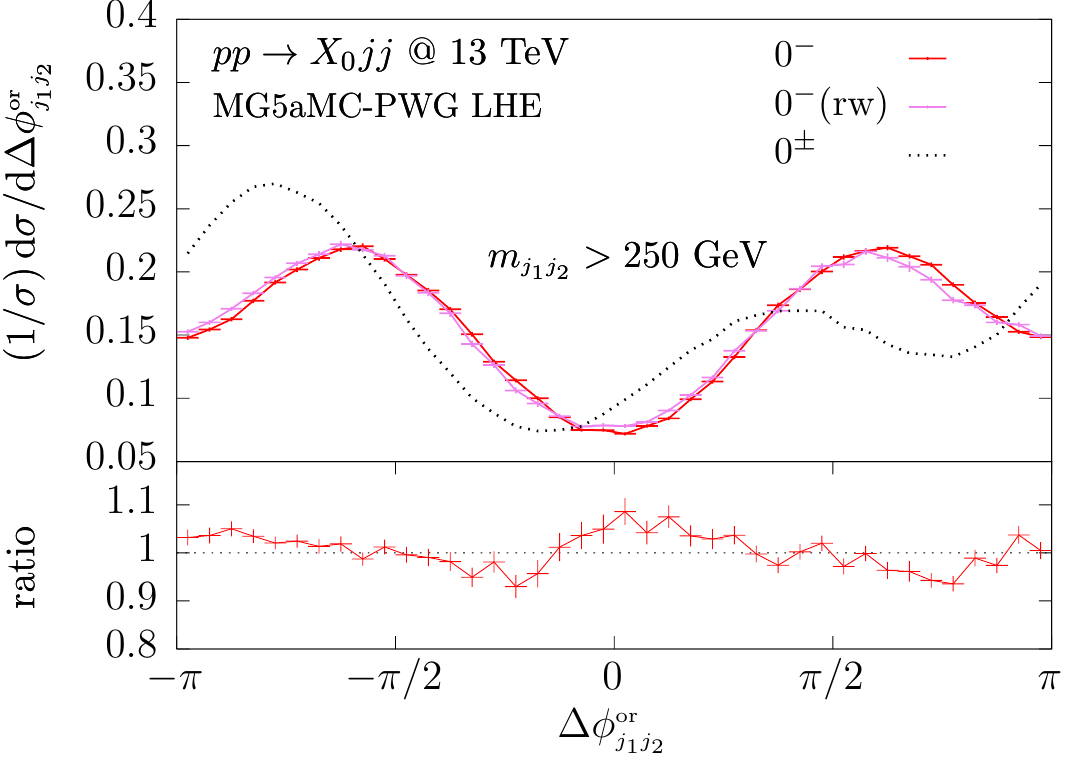}
  \end{center}
  \caption{Same as Fig.~\ref{fig:rwgt_from_scalar} but for the reweighting of
    the CP mixed sample to the scalar case (on the left) and to the
    pseudoscalar one (on the right).}
  \label{fig:rwgt_from_mixed}
\end{figure}

\begin{figure}[htb!]
  \begin{center}
    \includegraphics[width=0.7\textwidth]{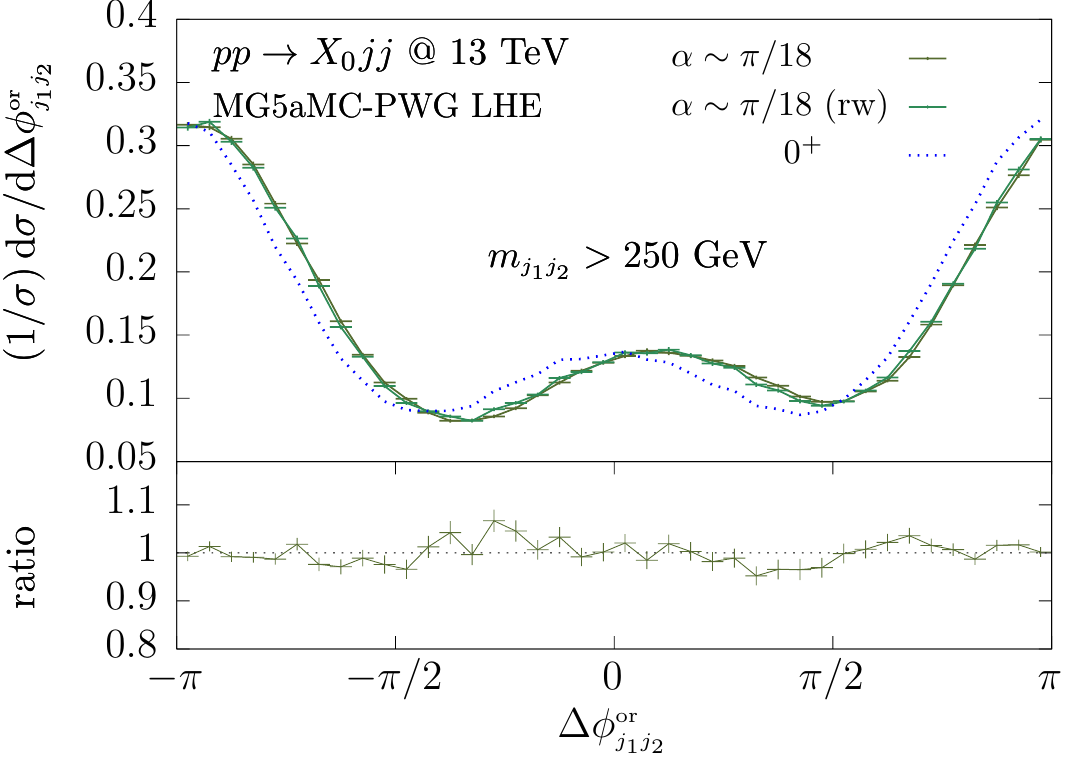}
  \end{center}
  \caption{Same as Fig.~\ref{fig:rwgt_from_scalar} but for the reweighting of
    the scalar sample to the CP scenario defined by $\cos\alpha=0.985$.}
   \label{fig:rwgt0985_from_scalar}
\end{figure}

In this section we present a few results obtained with the \pwgboxvtwo{}
reweighting feature.  We have reweighted two of the event samples that we
have produced: the scalar and the mixed one.  We have then compared the
reweighted distributions with the original ones, i.e.~those computed from the
beginning with a given value of $\cos\alpha$.  In particular, we have
reweighted the scalar sample to the pseudoscalar and CP mixed cases, and we
have reweighted the mixed sample to the scalar and pseudoscalar ones.  We
have found an overall good agreement between the reweighted and the original
distributions, except for the distribution of the differential cross section
expressed as a function of the oriented azimuthal angle, i.e.~the
distributions most sensitive to the value of the CP parameter $\cos\alpha$.

In Fig.~\ref{fig:rwgt_from_scalar} we compare three curves. The \xDphiorjj{}
distribution obtained from the original scalar sample is plotted in dotted
blue, on both panes. This curve corresponds to the $0^+$ line on the left
pane of Fig.~\ref{fig:deltaphij1j2or}.  The scalar sample is reweighted to
the pseudoscalar scenario on the left pane and to the mixed scenario on the
right pane.  The reweighted sample, indicated with ``rw'' in the figures, is
then compared with the original distribution.  The ratio of the last two
curves is also plotted.  In both cases, in correspondence to the minima of
the $0^+$ distribution, the discrepancy between the reweighted distribution
and the original one is more than $-10\%$, the minus sign to indicate that
the distributions obtained by reweighting underestimate the original ones.
The opposite is also true: when the $0^+$ distribution has maxima that are
not close to the maxima of the $0^-$ and $0^\pm$ distributions, we have a
discrepancy on the opposite side, up to $+10\%$.

Similar conclusions can be drawn by reweighting the $0^\pm$ sample, as
illustrated in Fig.~\ref{fig:rwgt_from_mixed}, in order to produce the
differential cross section as a function of \xDphiorjj{} for the $0^+$ and
$0^-$ scenarios.

These differences can be explained by noticing that the minima of the above
distributions are actually zeros at LO, and the production of events around
these regions is then suppressed. The reweighting procedure is not able to
generate the correct distributions, if the starting one is very different
from the final one, i.e., for example, going from $\alpha=0$ to
$\alpha=\pi/2$, for the reweighting of the scalar case to the pseudoscalar
one.

Otherwise, if the reweighting procedure is used to reweight distributions
with similar values of the angle $\alpha$, the procedure correctly works.
This is shown in Fig.~\ref{fig:rwgt0985_from_scalar}, where the distribution
computed with $\alpha=0$ is reweighted to the distribution with $\alpha\sim
10^\circ \sim \pi/18$, and the agreement with the exact one is very good.


\subsection{MiNLO}
In this section we present a few results for the pseudoscalar $X_0$
production, obtained within the MiNLO procedure.  Although all the cuts
applied on the jets in the previous sections are completely removed, the
differential cross sections for inclusive quantities are finite, due to the
presence of the MiNLO Sudakov form factor.

\begin{figure}[htb!]
  \begin{center}
    \includegraphics[width=0.495\textwidth]{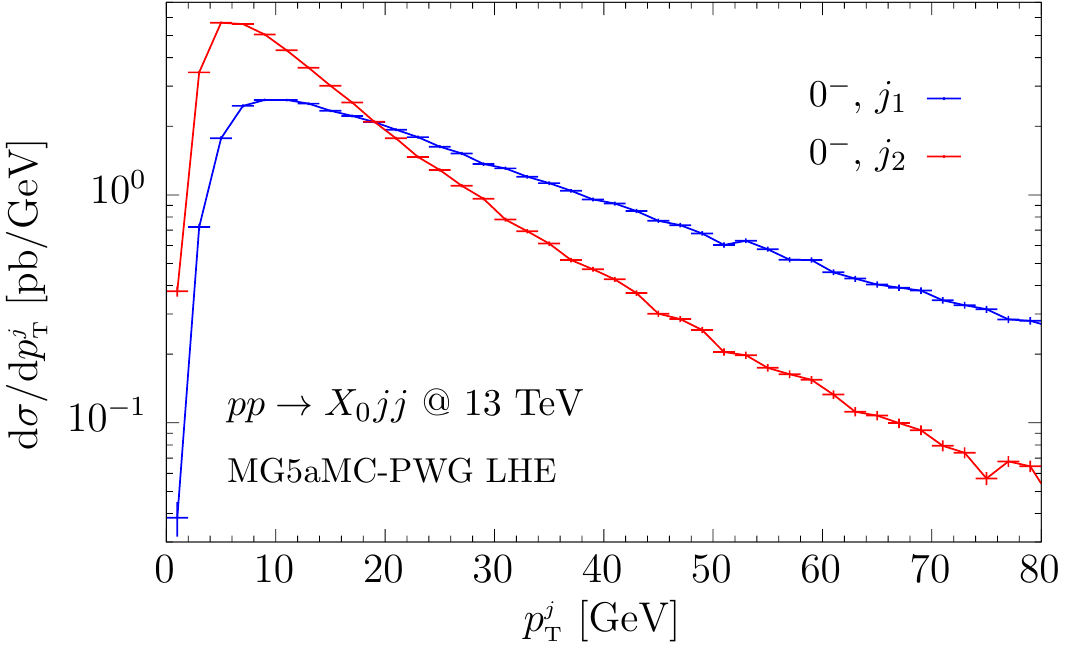}
    \includegraphics[width=0.495\textwidth]{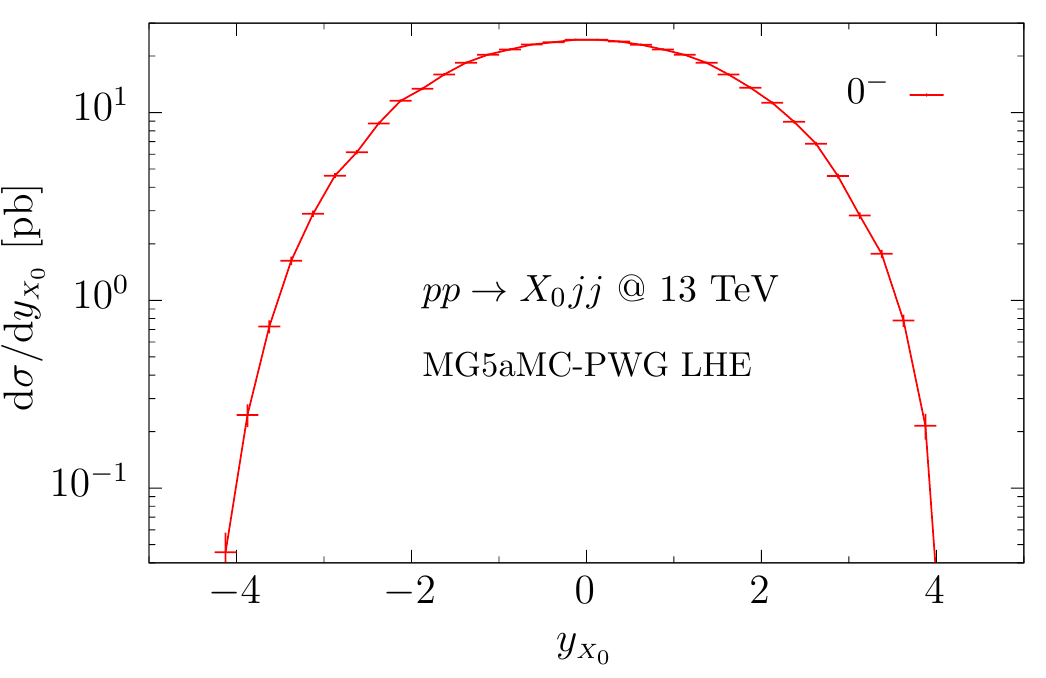}
  \end{center}
  \caption{On the left pane the inclusive differential cross section as a
    function of the transverse momentum of the hardest jet, in blue, and of
    the second-to-hardest one, in red. The CP scenario is defined by
    $\cos\alpha=0$, namely, the pseudoscalar case. On the right pane in red,
    the inclusive rapidity of the $X_0$ boson, for the same CP scenario as in
    the left pane. Both plots are obtained with MiNLO.}
  \label{fig:minlo}
\end{figure}

This is shown, for example, in Fig.~\ref{fig:minlo}, where we plot the
inclusive differential cross section as a function of the transverse momentum
of the hardest and of the second-to-hardest jet, on the left pane, and the
inclusive rapidity of the $X_0$ boson, on the right one.

Although finite, we cannot make any claim on the accuracy of these
distributions, i.e.~they do not reach the NLO accuracy of the MiNLO' method,
described in Refs.~\cite{Hamilton:2012rf, Frederix:2015fyz}.

\section{Conclusions}
\label{sec:conclusions}
In this paper we have presented an interface between \mglong{} and the
\pwgboxvtwo, able to build a NLO + parton shower generator for Standard Model
and many beyond Standard Model processes, in an automatic way.

The structure of the interface is such that future developments in \mglong{}
and \pwgboxvtwo{} remain independent to a large extent, so that it benefits
from all the progresses coming from both sides.
In fact, on the one side, \mglong{} provides the matrix elements for the Born,
the colour- and spin-correlated Born, the real and the virtual contributions.
On the other, the \pwgbox{} uses these ingredients to generate events
accurate at the NLO + parton shower level.
In addition, the interface writes other files needed by the \pwgboxvtwo{}.
Some of them, as the list of processes, are fully finalised.  Others, such as
the phase-space generator, need to be adjusted in order to deal with the
process at hand.

By now the interface only deals with processes for which we aim at NLO QCD
accuracy.  The extension including the electroweak corrections and the
interface with the more recent version of the \pwgbox{}, i.e.~the
\pwgboxres{}, is left as future work.

As a case study, using this interface we have generated the code for the
production of a spin-0 boson plus two jets, and we have computed a few
kinematic distributions, sensitive to the CP properties of the coupling of
the boson with a massive top quark. We have compared these distributions with
known results in the literature and found full agreement. We have also
presented a few results for the pseudoscalar case, obtained within the MiNLO
approach.

Finally, we have tested the \pwgbox{} reweighting feature. This procedure
works fine for every kinematic distributions we have examined, but for the
ones most sensitive to the CP nature of the $X_0$ boson. In fact, we have
observed that it works if the reweighting is done from one distribution to
another, with values of the mixing angle $\alpha$ not very different from
each other.


\section*{Acknowledgments}
P.N. acknowledges support from Fondazione Cariplo and Regione Lombardia,
grant 2017-2070, and from INFN.  We thank S.~Frixione and F.~Maltoni for
useful discussions.  We thank A.~Gritsan for suggesting the process we
implemented.


\providecommand{\href}[2]{#2}\begingroup\raggedright\endgroup

\end{document}